\documentclass[twocolumn,superscriptaddress,altaffilletter,prd]{revtex4-1}

\usepackage[utf8]{inputenc}
\usepackage{lineno,hyperref}
\usepackage{import}
\usepackage{amssymb}
\usepackage[italic]{heppennames}
\usepackage{fixltx2e}

\modulolinenumbers[5]

\usepackage[dvipsnames]{xcolor} 

\usepackage[draft]{pgf}

\usepackage{graphicx}
\usepackage{epstopdf}
\usepackage{color}
\usepackage{url}
\usepackage{soul}
\usepackage[normalem]{ulem}
\usepackage{amsmath}

\usepackage[flushleft]{threeparttable}
\usepackage{siunitx}

\usepackage{dcolumn}
\usepackage{bm}
\usepackage{color}
\usepackage{upgreek}

\usepackage{lineno}

\newcommand{\ndbd}{$ 0 {\nu} {\upbeta} {\upbeta} $}
\newcommand{\dbd}{$ 2 {\nu} {\upbeta} {\upbeta} $}
\newcommand{\isotope}[2]{$^{#2}$#1}
\newcommand{\sensitivityResult}{1.06$\times$10$^{26}$}
\newcommand{\sensitivityEnriched}{1.06$\times$10$^{27}$}
\newcommand{\nubar}{\skew3\bar\nu}

\begin{document}

\title{Projected sensitivity of the LUX-ZEPLIN experiment to the \texorpdfstring{\ndbd}~ decay of \isotope{Xe}{136}}


\author{D.S.~Akerib}
\affiliation{SLAC National Accelerator Laboratory, Menlo Park, CA 94025-7015, USA}
\affiliation{Kavli Institute for Particle Astrophysics and Cosmology, Stanford University, Stanford, CA  94305-4085 USA}

\author{C.W.~Akerlof}
\affiliation{University of Michigan, Randall Laboratory of Physics, Ann Arbor, MI 48109-1040, USA}

\author{A.~Alqahtani}
\affiliation{Brown University, Department of Physics, Providence, RI 02912-9037, USA}

\author{S.K.~Alsum}
\affiliation{University of Wisconsin-Madison, Department of Physics, Madison, WI 53706-1390, USA}

\author{T.J.~Anderson}
\affiliation{SLAC National Accelerator Laboratory, Menlo Park, CA 94025-7015, USA}
\affiliation{Kavli Institute for Particle Astrophysics and Cosmology, Stanford University, Stanford, CA  94305-4085 USA}

\author{N.~Angelides}
\affiliation{University College London (UCL), Department of Physics and Astronomy, London WC1E 6BT, UK}

\author{H.M.~Ara\'{u}jo}
\affiliation{Imperial College London, Physics Department, Blackett Laboratory, London SW7 2AZ, UK}

\author{J.E.~Armstrong}
\affiliation{University of Maryland, Department of Physics, College Park, MD 20742-4111, USA}

\author{M.~Arthurs}
\affiliation{University of Michigan, Randall Laboratory of Physics, Ann Arbor, MI 48109-1040, USA}

\author{X.~Bai}
\affiliation{South Dakota School of Mines and Technology, Rapid City, SD 57701-3901, USA}

% ...

\collaboration{The LUX-ZEPLIN (LZ) Collaboration}

\date{\today}

\begin{abstract}
The LUX-ZEPLIN (LZ) experiment will enable a neutrinoless double beta decay search in parallel to the main science goal of discovering dark matter particle interactions. We report the expected LZ sensitivity to \isotope{Xe}{136} neutrinoless double beta decay, taking advantage of the significant ($>$600~kg) \isotope{Xe}{136} mass contained within the active volume of LZ without isotopic enrichment. After 1000 live-days, the median exclusion sensitivity to the half-life of \isotope{Xe}{136} is projected to be \sensitivityResult~years (90\% confidence level), similar to existing constraints. We also report the expected sensitivity of a possible subsequent dedicated exposure using 90\% enrichment with \isotope{Xe}{136} at \sensitivityEnriched~years.
\end{abstract}

\maketitle


\section{INTRODUCTION}
\label{sec:intro}

Neutrinoless double beta decay (\ndbd) is a process by which a nucleus emits two electrons and no neutrinos. This is distinct from two-neutrino double beta decay (\dbd), whereby a nucleus emits two electrons and two electron antineutrinos ($\nubar_e$). \dbd~has been observed in several isotopes, and may be observable in even-even nuclei whenever single beta decay is energetically forbidden or highly suppressed \cite{Saakyan:2013}. In particular, \isotope{Xe}{136}, which comprises 8.9\% of naturally occurring xenon, has been shown to undergo \dbd~with a half-life of 2.165~$\pm$~0.016$^{\text{(stat)}}$~$\pm$~0.059$^{\text{(sys)}}$~$\times$~10$^{21}$ years with a Q-value of 2457.83~$\pm$~0.37~keV \cite{Albert:2013gpz, Redshaw:2007un}. There are thus far no unambiguous observations of \ndbd. This neutrinoless decay mode is allowed only if the neutrino is its own antiparticle, an idea originally suggested by Majorana \cite{Majorana:1937}. Particles which are their own antiparticles are referred to as Majorana particles. A \ndbd~decay would result in a mono-energetic peak in the $\upbeta$-spectrum at the double beta decay Q-value, Q$_{\beta\beta}$, because the electrons must carry almost all the energy of the decay (with a small fraction going into the recoiling nucleus), allowing one to separate the process from Standard Model \dbd. Observation of \ndbd~would imply the discovery of fundamental massive Majorana particles, lepton number violation (\textit{$\Delta$L} = 2), and \textit{B--L} violation (\textit{$\Delta$B} = 0 and \textit{$\Delta$L} = 2). Currently, the best upper limit on the half-life for \ndbd~of \isotope{Xe}{136} comes from the KamLAND-Zen experiment at 1.07$\times$10$^{26}$~years \cite{KamLAND-Zen:2016}.

A detector designed to observe the \ndbd~decay of a given source needs to have a complete understanding of the backgrounds in the event search region, a high abundance of the decaying element to compensate for the rare nature of this process and an excellent energy resolution at the Q-value of the decay.

The LUX-ZEPLIN (LZ) experiment features a two-phase xenon time projection chamber (TPC) designed to search for weakly interacting massive particles (WIMPs). While WIMPs present an entirely different signal than \ndbd, many of the experimental challenges are similar. Both require low backgrounds, a large active mass and, in the case of a xenon TPC, good scintillation and ionization collection. Most \ndbd~experiments use sources enriched in the isotope of interest, so as to increase the fraction of the relevant isotope and to decrease the passive non-source material. LZ's active detection mass is 7 tonnes of natural liquid xenon (LXe), yielding 623~kg of \isotope{Xe}{136} at natural abundance. This is a comparable mass to other world-leading \ndbd~experiments. Here the sensitivity of LZ to \ndbd~is investigated and compared to current limits and other next generation \ndbd~searches using xenon, including nEXO, NEXT, KamLAND2-Zen, and PandaX-III \cite{Albert:2017hjq,NEXT,KamLAND-Zen:2016,Chen:2016qcd}.

\section{THE LZ DETECTOR}
\label{sec:LZoverview}

\begin{figure*}[ht]
\includegraphics[width=0.65\textwidth]{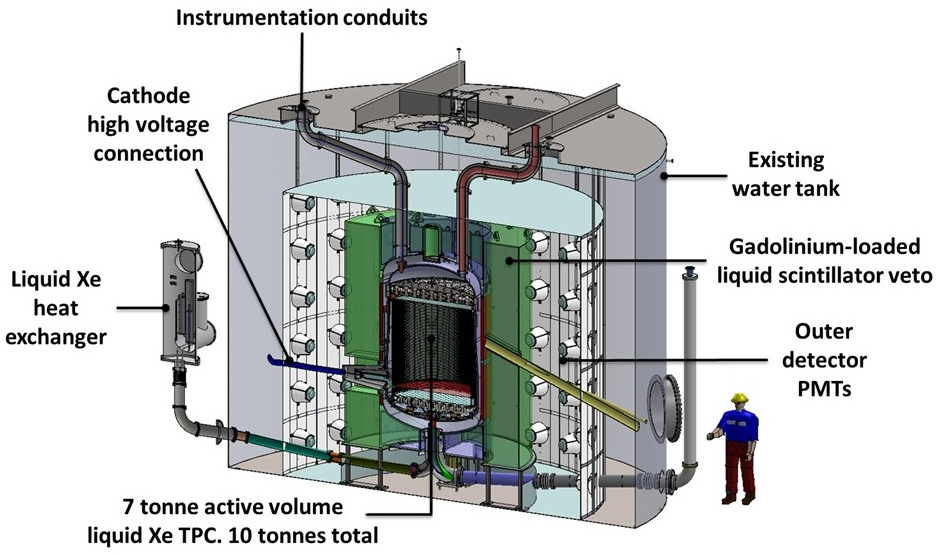}
\caption{\label{LZSolid_Jan16} Schematic of the LZ detector: an outer water tank shields external radioactivity. An outer detector system consisting of a Gd-doped liquid scintillator in acrylic designed to tag outgoing neutrons and gammas is indicated in green. The outer detector is viewed by PMTs to tag energy deposits in the scintillator as well as muons passing through the water tank. The TPC sits inside a titanium cryostat surrounded by the outer detector acrylic tanks. Most of the xenon is contained inside the inner cryostat vessel.}
\end{figure*}

A schematic of LZ is shown in Figure \ref{LZSolid_Jan16}. LZ will occupy the Davis cavern at the Sanford Underground Research Facility (SURF) in Lead, South Dakota (USA) in the location where the LUX experiment operated from 2012 until 2016 \cite{lab1,lab2,lab3,lab3-2,lab4}. In a two-phase xenon TPC such as LZ, energy deposits produce prompt scintillation light (S1) and free electrons. Some electrons recombine with the xenon ions producing more scintillation light. The electrons that do not recombine drift in an electric field to the liquid surface where they are extracted into a high-field gas region, creating a proportional scintillation signal (S2) \cite{Dolgoshein:1970,Benetti:1993,Chepel:2012sj}. In LZ both signals are detected by two arrays of Hamamatsu R11410-22 3-inch diameter low-radioactivity photomultiplier tubes (PMTs) \cite{LUXPMTPaper} that observe the active region from the top and bottom: 253 in the top array and 241 in the bottom array.
The relative intensity of S2 light in each PMT is used to reconstruct the event position in the horizontal ($x$,$y$) plane. The time difference between the S1 and S2 pulses indicates the free electron drift distance and is used to determine the depth ($z$) of the interaction. The drift field is created by the voltage difference between a cathode grid at the bottom of the detector and a gate grid just below the liquid surface. The electron extraction field around the liquid-gas interface is produced by the voltage applied between the gate grid and the anode grid above the LXe volume. The TPC has a drift region of 145.6~cm (from cathode to gate grid) and an inner diameter of 145.6~cm, and contains 7 tonnes of active xenon. A bottom grid situated below the cathode prevents the bottom PMT array from being exposed to the high fields near the cathode grid. The region between the bottom grid and the cathode grid is 13.75~cm in depth and is referred to as the reverse field region \cite{LZTDR:2016}.

 The TPC is contained inside a low-background, double-walled titanium cryostat \cite{LZTi:2016}, containing approximately 10 tonnes of liquid xenon. The TPC is surrounded by an additional volume of xenon instrumented with PMTs, referred to as the xenon ``skin''. The side section of the skin that surrounds the TPC is 4~cm thick at the top and 8~cm at the level of the cathode. The total amount of xenon in the full skin system is around 2 tonnes. Light collection efficiency of the skin detector is highly dependent on the position of the interaction. Studies of PMT coverage and wall reflectivity have led to an expected gamma-ray energy threshold  of 100 keV in more than 95\% of the skin region.

The cryostat vessel is surrounded by the Outer Detector (OD), containing organic liquid scintillator (gadolinium-loaded linear alkylbenzene), and a water shield. These systems will be viewed by an array of 120 8-inch diameter Hamamatsu R5912 PMTs, providing an additional active veto for gamma-ray scatters in the liquid scintillator. 
The average light collection efficiency over the entire outer detector volume is estimated to be $\sim$7\%, providing an average light yield of about 130 photoelectrons for a 1~MeV energy deposit in the liquid scintillator \cite{LZTDR:2016}. The outer detector system records PMT data in tandem with a triggered event in the LXe TPC, meaning that the energy threshold of the outer detector is only limited by the light collection and the background rate from \isotope{C}{14} decay in the liquid scintillator (Q-value of 156~keV). The prompt coincidence window for the outer detector veto is set as 1\,$\upmu$s for this analysis to reduce the rate of accidental coincidences due to \isotope{C}{14} decays in coincidence with a potential \ndbd~event in the TPC. This coincidence window can be much smaller than that considered for the main WIMP search analysis (500~$\upmu$s) since neutron interactions are not relevant at the \ndbd~energies, allowing the threshold of the outer detector to be lowered from 200~keV to 100~keV without increasing the dead time significantly. A more detailed description of the LZ experiment can be found in \cite{LZTDR:2016}.

\subsection{Event Reconstruction and Energy Resolution}
The energy of an interaction in the forward field region of the TPC can be estimated by combining both the S1 and S2 signals through the relation, 
\begin{equation}
    E = W \left(\frac{\text{S1}}{\textit{g}1} + \frac{\text{S2}}{\textit{g}2}\right),
    \label{eq:S1S2energy}
\end{equation}
where \textit{W} = 13.7 eV is the \textit{W}-value of liquid xenon \cite{PLATZMAN1961116} and \textit{g}1, \textit{g}2 are detector dependent gain factors. \textit{g}1 corresponds to the product of the detector light collection efficiency and PMT quantum efficiencies and \textit{g}2 is the product of the efficiency to extract electrons to the gas phase and the pulse size of a single electron signal. These factors are determined directly from data using calibration sources of known energies.

Because the S1 and S2 signals are proportional to the number of scintillation photons and free electrons, each fluctuates due to effects of recombination. Some of the free electrons may recombine, resulting in fewer free electrons and more photons. In this way, for a scatter of fixed energy, there is an anti-correlation between the S1 and S2 signals. Therefore, combining the S1 and the S2 signals to measure deposited energy will substantially improve the energy resolution \cite{Conti:2003av} beyond what may be achieved with the S1 or S2 signals individually.

As the range of energies for a \ndbd~search is much higher than for dark matter searches, it is expected that some PMT saturation might occur for the S2 signal. A combination of simulation results combined with the PMT voltage divider design indicate that a \ndbd~S1 will not saturate the PMTs, nor would its S2 saturate the PMTs in the bottom array. However, some PMTs in the top array will saturate the S2 signal due to limited dynamic range of the ADC and/or to capacitor depletion in the PMT biasing circuit \cite{FahamThesis}. Around 7 to 22 PMTs in the top array will saturate the ADCs depending on the drift time and the resulting level of electron diffusion before S2 production. Nonlinearity effects caused by capacitor depletion, which affects only the top PMTs in the S2 signal, will be corrected during data processing. For an S2 signal, approximately 79 photons detected (phd) per extracted electron are assumed with 95\% extraction efficiency, with 27 of those photons being detected in the bottom array \cite{PhysRevD.101.052002}. As such, the bottom array alone can be used to estimate the S2 size with minimal impact on the energy resolution. However, the top array is used to reconstruct the $xy$ position of events, so the effects of saturation on $xy$ position reconstruction must be considered. The event position is useful for rejecting radioactive backgrounds which are higher in rate at the edge of the detector. The effects of saturation can be minimized by excluding the saturated PMTs from the position reconstruction \cite{Akerib:2017riv}. Preliminary studies demonstrate that LZ will achieve a $xy$ position resolution of 0.5~cm or better for interaction with deposited energies above 1.8~MeV within a radial distance of 68.8~cm from the center of the detector and across the full drift length of the TPC. Near the center of the detector, the resolution will be 0.2~cm or better.

\subsection{Detector Calibration}

The self-shielding provided by liquid xenon makes detector calibration challenging. For this reason, LZ utilizes several radioactive sources that can be injected into the active xenon, which can be removed later or decay away with a short half-life \cite{Akerib:2015wdi,Akerib:2017vbi}. However, almost all of these sources deposit less than 200~keV in one interaction, making calibration at higher energies more difficult. One option is to use a \isotope{Rn}{220}~source as suggested by \cite{Lang:2016}. The energy spectrum of the \isotope{Tl}{208}~daughter has a step around 3.2~MeV created by the coincident beta and gamma decays, and also has alpha decays at 6.2, 6.4, 6.9, and 8.9~MeV. A \isotope{Rn}{220}~calibration is already planned for LZ, as well as an external \isotope{Th}{228}~gamma-ray source that will, at the very least, calibrate the outer regions of the active xenon. The internal sources allow for robust correction for the position dependence of the detector response, so performing a high-energy ($\sim$MeV) calibration uniformly through the xenon is not necessary. Lastly, although the backgrounds are very low in the inner volume of the TPC, there is a plethora of visible gamma-ray lines in the outside sections of the active region. Although one would not want to rely on these as the only detector calibration, the LZ background data stands as a useful crosscheck on any high-energy calibration.

\section{BACKGROUND MODEL}
\label{sec:backgrounds}

The main background contributions for this \ndbd~search are summarized in Table \ref{tab:BG}. Extensive Monte Carlo simulations of the backgrounds due to radioactive contamination in detector components and the cavern rock are generated using BACCARAT, a framework based on GEANT4 that evolved from the LUXSim \cite{LUXSim} simulation package which is also used to make predictions for dark matter sensitivity \cite{PhysRevD.101.052002}. The full detector geometry is modeled and matches the engineering drawings. The statistics generated for each simulated component typically corresponds to tens to hundreds of thousands of days, significantly exceeding the run time of the experiment, which is considered here to be 1000 days. 
The model used in this analysis was constructed using the most recent material assays and detector simulations \cite{PhysRevD.101.052002,LZassaysPaper:prep}. The uncertainty in the estimated backgrounds is dominated by the uncertainties in the cavern rock gammas flux measurements (50\% for \isotope{U}{238} and 25\% for \isotope{Th}{232}) \cite{lab5}. However, a 1$\sigma$ increase in the total rock gamma background would only result in a 5\% decrease of the sensitivity. Backgrounds will be measured with high accuracy once the detector begins taking data. The outer detector and skin systems will be integral for the characterization of backgrounds, specially those from external sources like gammas from cavern rock. Events recorded near the LXe wall can also be used to describe and constrain several backgrounds from both internal and external sources. All data will be fit to simulations in order to best constrain the different backgrounds observed.

\begin{table*}[t]
  \begin{threeparttable}
  \caption{\label{tab:BG}Summary table of the masses, activities and estimated background counts in the $\pm~1\sigma$ ROI and inner 967~kg mass, for a 1000 day run, considering 1.0\% energy resolution at Q-value and 0.3~cm multiple scatter rejection along $z$ (see text for details).}
\begin{tabular}{ lcccccr}
\hline
 Item & \multicolumn{1}{c}{Mass} & \multicolumn{1}{c}{\isotope{U}{238}-late} &  \multicolumn{1}{c}{Counts} & \multicolumn{1}{c}{\isotope{Th}{232}-late} & \multicolumn{1}{c}{Counts} & \multicolumn{1}{c}{Total } \\ 
    & \multicolumn{1}{c}{(kg)} & \multicolumn{1}{c}{(mBq/kg)} &  \multicolumn{1}{c}{from \textsuperscript{238}U} & \multicolumn{1}{c}{(mBq/kg)} & \multicolumn{1}{c}{from \textsuperscript{232}Th} & \multicolumn{1}{c}{ Counts} \\ 
  \hline \hline
  TPC PMTs & 91.9 & 3.22 & 2.95 & 1.61 & 0.10 & 3.05 \\
  TPC PMT bases & 2.80 & 75.9 & 1.52 & 33.1 & 0.03 & 1.55 \\
  TPC PMT structures & 166 & 1.60 & 2.65 & 1.06 & 0.12 & 2.77 \\
  TPC PMT cables & 88.7 & 4.31 & 1.44 & 0.82 & 0.19 & 1.63 \\
  Skin PMTs and bases & 8.59 & 46.0 & 0.75 & 14.9 & 0.02 & 0.78 \\
  PTFE walls & 184 & 0.04 & 0.39 & 0.01 & 0.00 & 0.39 \\
  TPC sensors & 5.02 & 5.82 & 1.19 & 1.88 & 0.00 & 1.19 \\
  Field grids and holders & 89.1 & 2.63 & 0.62 & 1.46 & 0.11 & 0.73 \\
  Field-cage resistors & 0.06 & 1350 & 2.63 & 2010 & 0.03 & 2.65 \\
Field-cage rings & 93.0 & 0.35$^{\dagger}$ & 0.82 & 0.24$^{\dagger}$ & 0.00 & 0.82 \\
  Ti cryostat vessel & 2590 & 0.08$^{\dagger}$ & 1.30 & 0.22$^{\dagger}$ & 0.20 & 1.49 \\
  Cryostat insulation & 13.8 & 11.1$^{\dagger}$ & 0.90 & 7.79$^{\dagger}$ & 0.04 & 0.94 \\
  Outer detector system & 22900 & 4.71$^{\dagger}$ & 1.70 & 3.73$^{\dagger}$ & 1.08 & 2.79 \\
  Other components & 438 & 1.83 & 2.10 & 1.65 & 0.31 & 2.41 \\
  \hline
  Det. components subtotal & \multicolumn{1}{c}{-} &\multicolumn{1}{c}{-} & 21.0 & \multicolumn{1}{c}{-} & 2.32 & 23.3 \\
  Cavern walls & \multicolumn{1}{c}{-} & 29000.00 & 3.21 & 12500.00 & 8.41 & 11.6 \\
  Neutron-induced \isotope{Xe}{137} & \multicolumn{1}{c}{-} &\multicolumn{1}{c}{-} & \multicolumn{1}{c}{-} & \multicolumn{1}{c}{-} & \multicolumn{1}{c}{-} & 0.28$^{*}$ \\
  Internal \isotope{Rn}{222} & \multicolumn{1}{c}{-} & \multicolumn{1}{c}{-} & \multicolumn{1}{c}{-} & \multicolumn{1}{c}{-} & \multicolumn{1}{c}{-} & 0.45$^{*}$ \\ 
  \isotope{Xe}{136} \dbd & \multicolumn{1}{c}{-} & \multicolumn{1}{c}{-} & \multicolumn{1}{c}{-} & \multicolumn{1}{c}{-} & \multicolumn{1}{c}{-} & 0.01$^{\dagger}$ \\
  \isotope{B}{8}~solar neutrinos & \multicolumn{1}{c}{-} & \multicolumn{1}{c}{-} & \multicolumn{1}{c}{-} & \multicolumn{1}{c}{-} & \multicolumn{1}{c}{-} & 0.03 \\\hline \hline
  Total & \multicolumn{1}{c}{-} & \multicolumn{1}{c}{-} & 24.2 & \multicolumn{1}{c}{-} & 10.7 & 35.6 \\
\end{tabular}
\begin{tablenotes}
      \small
      \item \textit{$^{\dagger}$Upper limit}
      \item \textit{$^{*}$preliminary estimate}
      \item \textit{\isotope{U}{238}-late chain is \isotope{Ra}{226}~and after. \isotope{Th}{232}-late chain is \isotope{Ra}{224}~and after}
    \end{tablenotes}
  \end{threeparttable}
\end{table*}

There are two important gamma lines to consider for backgrounds near Q$_{\beta\beta}$. There is a line at 2614.5~keV from \isotope{Tl}{208}~decay in the \isotope{Th}{232}~decay chain, which is about 160~keV higher in energy than Q$_{\beta\beta}$ and has a branching ratio of 35.9\%. The second, more problematic, gamma line is from \isotope{Bi}{214}~ (\isotope{U}{238}-chain) at 2447.7~keV with 1.5\% branching ratio. This low branching ratio is fortunate, as this line cannot be separated from the signal with the energy resolution of LZ. There is also the possibility of a sum peak from \isotope{Co}{60}~at 1173.2\,+\,1332.5\,=\,2505.7~keV. However, simulations show that good rejection of multiple scatter events will eliminate this background.

An inner volume was defined with the goal of characterizing the relevant backgrounds for this analysis. This inner volume was optimized using a cut-and-count analysis and provides a snapshot of the full background model in the most sensitive region of the detector. This volume is defined within 26~$<z<$~96~cm and for radii smaller than 39~cm, containing $\sim$967~kg of LXe. A larger fiducial volume is used for the sensitivity analysis, as discussed in Section \ref{sec:sensitivity}. The region-of-interest (ROI) considered on this analysis is 2433.3~$<E_{dep}<$~2482.4~keV, representing a $\pm~1\sigma$ energy window around Q$_{\beta\beta}$, considering an energy resolution ($\sigma$/E) of 1\% (see Section \ref{subsec:DetPerformance}). This energy window is used to characterize the backgrounds in the central \ndbd~signal region. The sensitivity analysis uses an extended range of energies, from 2000~keV to 2700~keV, in order to model the backgrounds more precisely, as discussed in Section \ref{sec:sensitivity}.

\begin{figure}[ht]
\includegraphics[width=\linewidth]{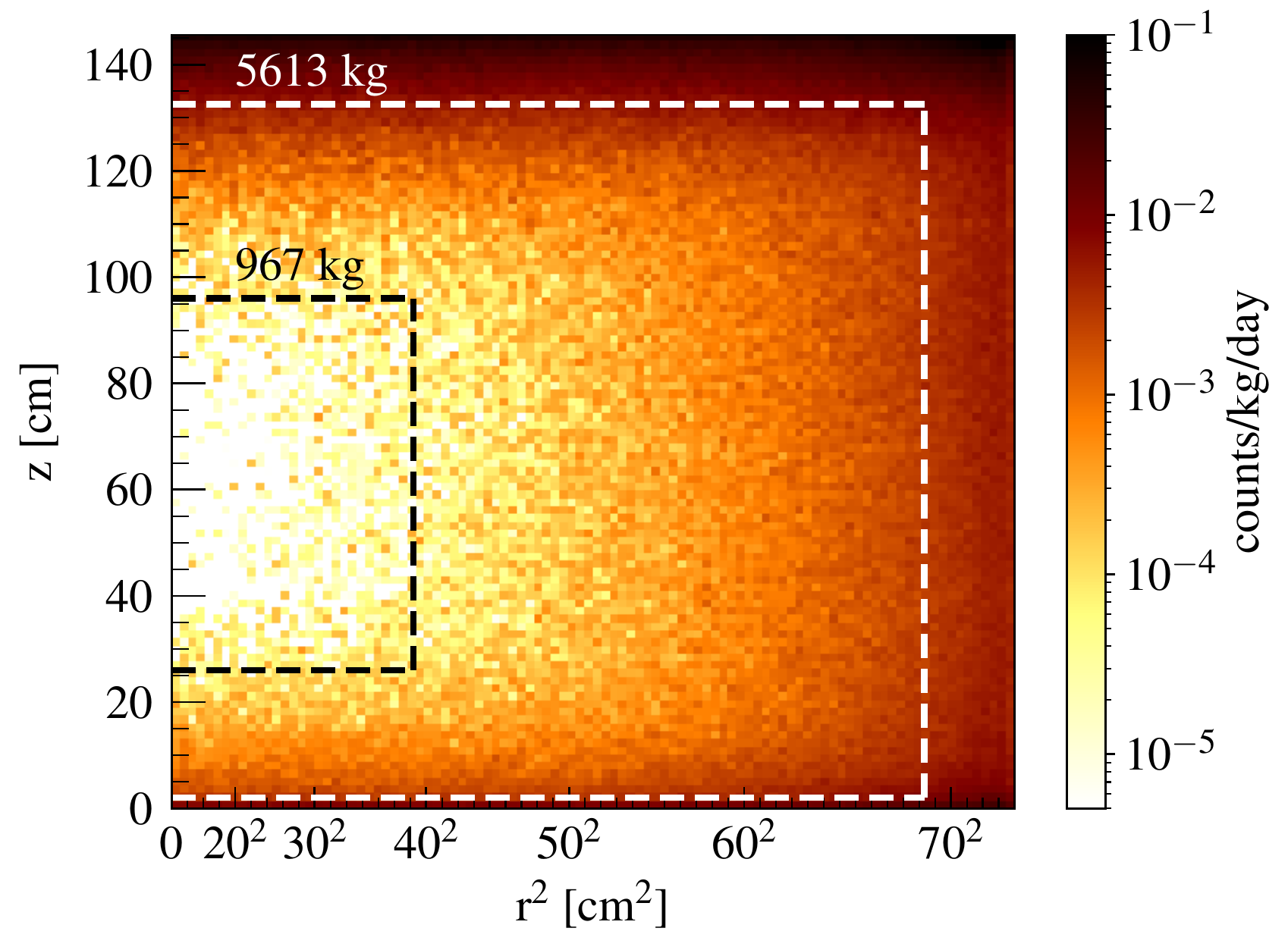}
\caption{Background event rate in the active region and in the $\pm~1\sigma$ energy ROI as a function of $r^2$ and $z$. The dashed black rectangle represents the inner 967~kg volume where LZ is most sensitive to the \ndbd~decay, while the larger dashed white rectangle represents the extended fiducial volume used on the profile likelihood analysis to obtain the LZ sensitivity to the \isotope{Xe}{136} \ndbd~decay.}
\label{fig:fiducialization}
\end{figure}

\begin{figure*}[ht]
\includegraphics[width=0.49\linewidth]{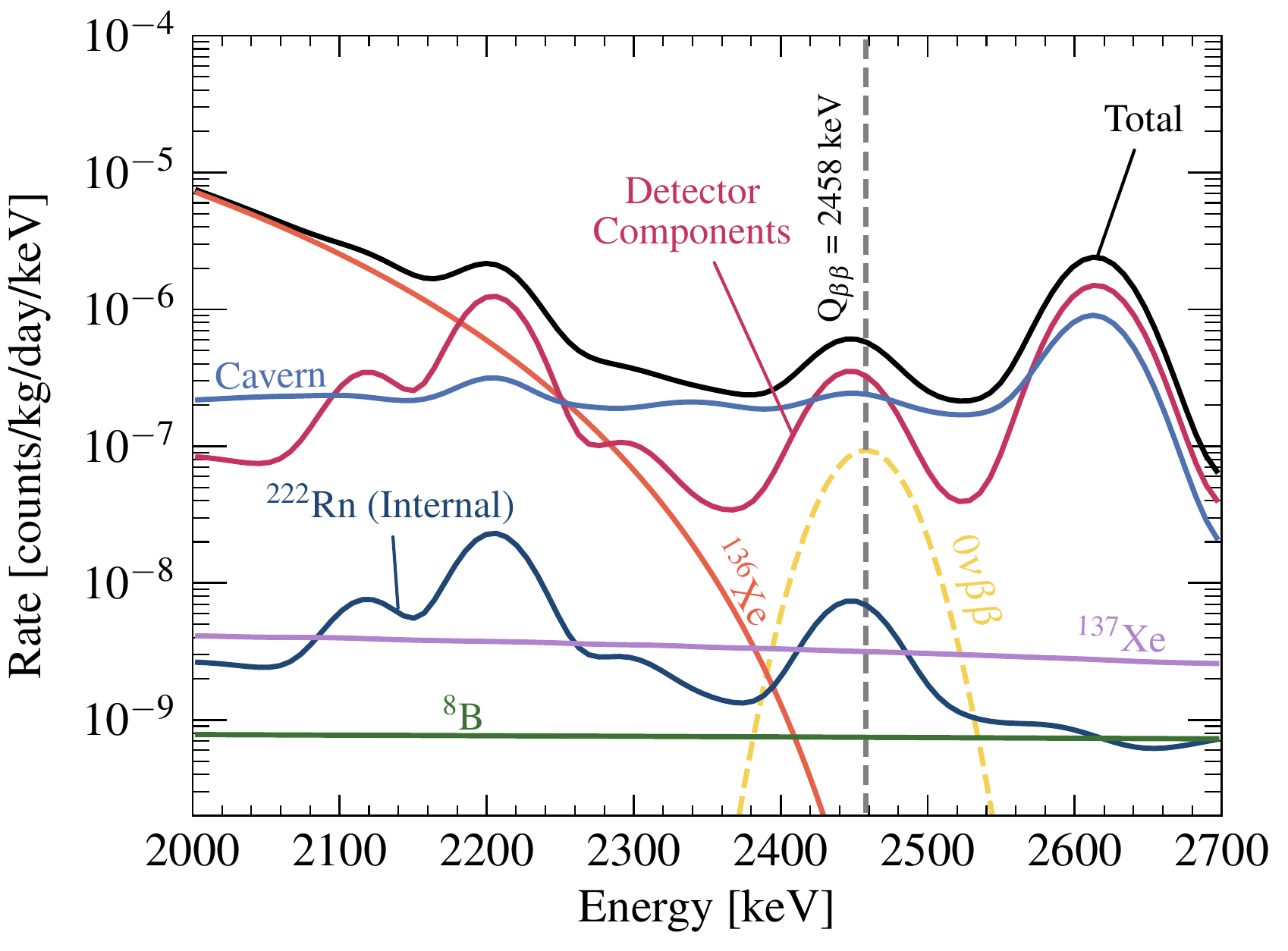}
\includegraphics[width=0.49\linewidth]{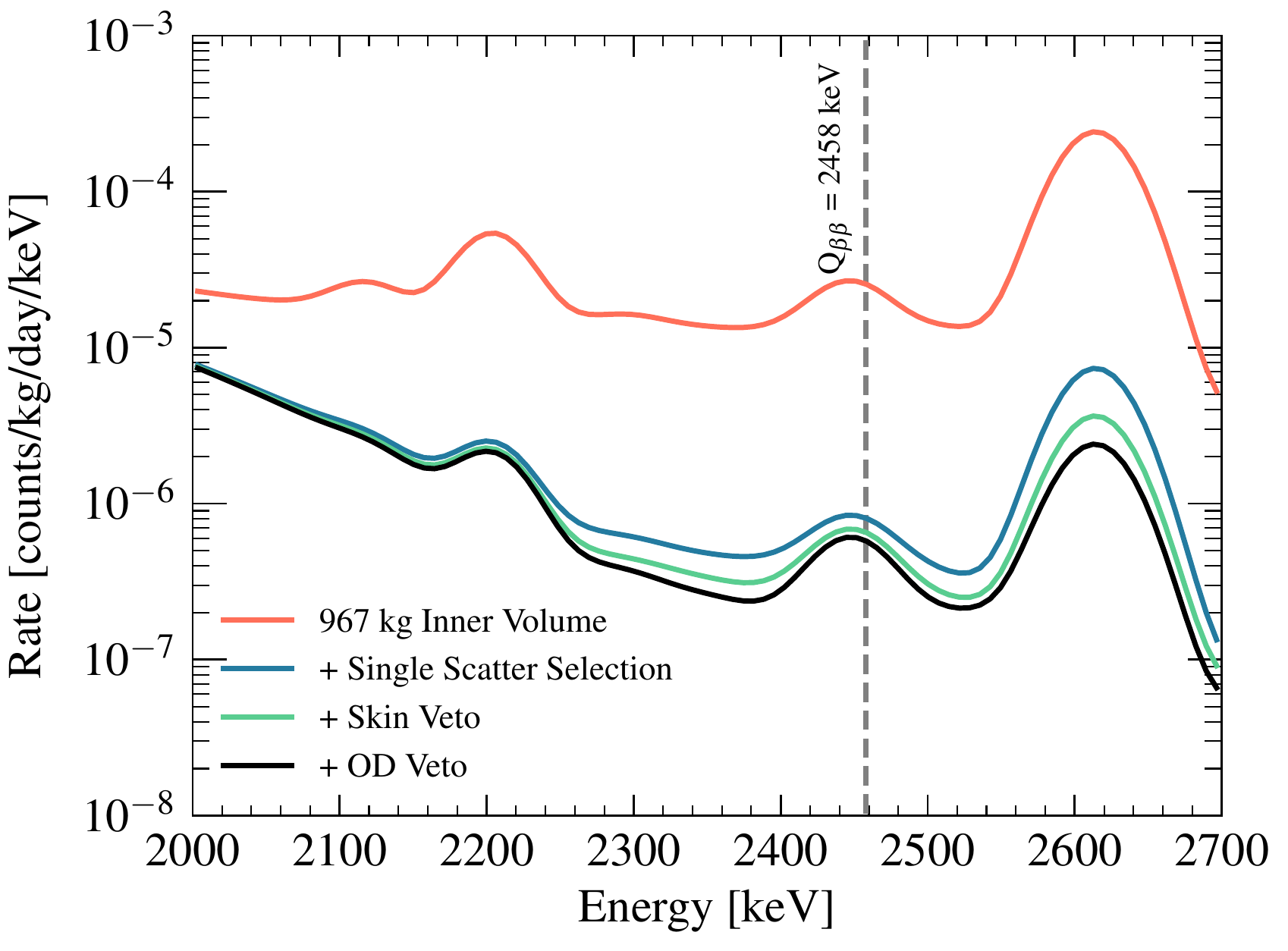}
\caption{Background energy spectrum in the inner 967~kg volume: contributions from the main background sources (left) and effect of successive selection cuts used in the analysis (right). More details regarding the analysis can be found in the text. The \isotope{Co}{60}, \isotope{U}{238}-late chain and \isotope{Th}{232}-late chain backgrounds from the detector components are combined into a single curve on the left plot but are treated independently in the sensitivity analysis. The dashed yellow line in the left plot represents the expected signal spectrum for \isotope{Xe}{136}~\ndbd~decay, considering a half-life of \sensitivityResult~years (see Section \ref{sec:sensitivity}), and it is not included in the total spectrum. The spectra are smeared using the energy resolution function of LUX \cite{LUXER:2016}, scaled to be $1.0\%$ at Q$_{\beta\beta}$.}
\label{fig:ESpectrum}
\end{figure*}

Figure \ref{fig:fiducialization} shows the number of simulated background events in the ROI versus $z$ and radius squared. The background rates are higher at the top than at the bottom of the active volume, as the bottom PMTs are shielded by the xenon in the reverse field region. The innermost region of the detector has a much lower background due to the self shielding of LXe. 
Figure \ref{fig:ESpectrum} on the left shows the background spectrum for the major contributors, as well as the total background spectrum, for a run lasting 1000 live-days and within the inner 967~kg volume. The right-hand side plot of Figure \ref{fig:ESpectrum} displays how the successive selection cuts used in this analysis impact the background spectrum in the inner 967~kg volume. A detailed explanation of the selection criteria used in the analysis can be found in Section \ref{sec:sensitivity}. The ``single scatter'' selection provides the strongest background rejection for gammas of these energies. However, this analysis cut does not exclude events from the \dbd~decay of \isotope{Xe}{136}, resulting in the loss of rejection efficiency visible at lower energies on Figure \ref{fig:ESpectrum}.

\subsection{Assumptions About Detector Performance}
\label{subsec:DetPerformance}

The energy resolution at the Q-value affects the experiment's ability to reject backgrounds from the 2614.5~keV \isotope{Tl}{208}~line. Using baseline assumptions about light collection (7.5\% photon detection efficiency averaged over the active volume), electron extraction efficiency (95\%), and single electron amplification (79 detected photons per extracted e$^{-}$) for LZ and applying the most recent NEST model \cite{NEST1,NEST2}, an energy resolution of 0.88\% at Q$_{\beta\beta}$ is predicted to be achievable with LZ. 
XENON1T demonstrates that an energy resolution of 0.79\% at Q$_{\beta\beta}$ is achievable with a tonne-scale dual-phase xenon detector \cite{Aprile:2020yad}. The drift field of XENON1T is 80 V/cm, significantly lower than the expected 310 V/cm drift field of LZ \cite{PhysRevD.101.052002} and the energy resolution is expected to improve with higher drift field as observed by EXO-200 \cite{Anton:2019hnw}. An energy resolution of 1.0\% at Q$_{\beta\beta}$ is assumed for this analysis, as a conservative value that will likely be improved.


Sensitivity estimates also depend on the minimal vertex separation needed to identify a multiple scatter event. Previous work assumed multiple scatter events could be rejected down to 3~mm separations in $z$ \cite{Baudis:2013qla}, and the same is assumed here.  
In LZ, the full-width at half-maximum of an S2 from a shallow event will be $\sim$1~$\upmu$s and a drift velocity $\leq$ 2~mm/$\upmu$s, so in theory separating scatters 2~mm apart should be possible. However, rejecting events with vertex separation in depth less than 3~mm would lower the efficiency for detecting \ndbd. Diffusion will widen the S2 pulses; this is a known and well-understood Gaussian smearing and can be addressed through advanced analysis, such as applying a drift-time dependent S2 width cut or pulse deconvolution. 
Vertex separation in the $xy$ plane is not used to reject multiple scatter events in this analysis. Preliminary studies show that a vertex separation in the $xy$ plane of 3~cm would only improve background rejection by less than 4\% in the inner 967~kg volume.

Finally, any event which deposits more than 100~keV in the LXe skin and/or 100~keV in the outer detector is vetoed.
This provides an advantage in tagging the 2614.5~keV \isotope{Tl}{208}~line, which is always emitted with another gamma-ray of at least 583~keV. The skin and outer detector are less helpful for reducing the background from the \isotope{Bi}{214}~line.

\subsection{Detector Components}


Every component of the TPC, skin, outer detector and auxiliary systems of LZ is included in the background model. Their individual contributions are estimated using the dedicated BACCARAT simulations mentioned above. Table \ref{tab:BG} summarizes the activities of these components. The categories \isotope{U}{238}-late and \isotope{Th}{232}-late refer only to the late chain activity, beyond \isotope{Th}{230} and \isotope{Th}{228}, respectively, as several samples measured by LZ were found to be out of secular equilibrium \cite{LZTDR:2016}.

All constituent materials have been screened for radioactivity directly by the LZ collaboration in specially designed facilities, with the exception of a few that are based on assay results from previous experiments \cite{LZassaysPaper:prep}. Some contamination values for detector materials are measured upper limits. Material assays will continue with the goal of improving the upper limits on several components. 

The TPC PMTs are the largest contributor to the backgrounds amongst detector components. The titanium vessels, with a mass of 2.6 tonnes, contributes only half of the total background counts from the full cryostat. Despite the low contamination levels of the outer detector acrylic tanks, the total acrylic mass of 4.3 tonnes is mostly responsible for the background rates from that system. The opposite is true for the resistors, which weigh less than 60 grams combined but have high contamination levels. Field-cage and cathode feedthrough resistors use interconnects while for other electronics components low activity solders were used \cite{LZassaysPaper:prep}.


\subsection{Davis Cavern Walls}

The rock surrounding the Davis cavern is composed primarily of amphibolite and rhyolite \cite{lab4,lab5} that was sprayed with a layer of shotcrete with an average thickness of 12~cm.
Recent measurements with a sodium iodide detector in the Davis cavern indicate an average activity of 12.5~Bq/kg of \isotope{Th}{232}~and of 29.0~Bq/kg of \isotope{U}{238}~in the surrounding cavern rock  \cite{lab5}. Measurements of gravel from beneath the water tank have shown an activity of 1.7~Bq/kg of \isotope{Th}{232}~and of 26.3~Bq/kg of \isotope{U}{238}~\cite{lab5}. LZ will be shielded from external gamma-rays by a 7.62~m diameter water tank, as well as 6 octagonal steel plates of 5 cm thickness embeded between the concrete and the bottom of the water tank. Although this shielding is more than sufficient for dark matter data taking, an 8~cm thickness of additional steel shielding above the water tank was added to enable a competitive double beta decay analysis. No additional shielding is assumed to be on the sides of the water tank. The outer detector system will be an important tool to mitigate the effect of the gammas from the laboratory rock. In order to tag events from the 2614.5~keV line from \isotope{Tl}{208}, which is only 160~keV away from the Q-value of the \dbd, the outer detector energy threshold is set at 100~keV for this analysis. The outer detector vessels will contain liquid scintillator with a thickness above 60~cm on top, bottom and sides (see Figure \ref{LZSolid_Jan16}).

Despite the shielding provided by the several steel plates, scintillator and the water tank, the background from rock gammas is significant. The contributions from the full \isotope{U}{238} and \isotope{Th}{232} chains were estimated using a detailed set of simulations that included the top and bottom steel plates and both veto systems. The effects of the shielding provided by the top steel plate, water tank, veto systems and detector materials results in a limited number of events reaching the active region and a subsequent loss of efficiency for simulations of sources in the rock. For that reason the same event biasing technique was used as in \cite{LZTDR:2016,lab5,Akerib:2020ewf}.

The simulations of gammas from the rock predict a non-negligible background rate from high energy gammas with energies above 3~MeV. Both the \isotope{U}{238} and \isotope{Th}{232} decay chains have several $\upalpha$ emissions with energies up to 8.8~MeV that may produce these high energy gammas in $(\alpha,\gamma)$ reactions in the rock on oxygen and silicon \cite{GammasSUL}, the most abundant elements in the cavern rock at SURF. Fortunately, the expected flux of these high energy gammas is 2 to 3 orders of magnitude lower than the radiogenic gammas from the \isotope{U}{238} and \isotope{Th}{232} decay chains, resulting in a total contribution to the background of less than one count in the inner 967~kg volume and in 1000 live-days, despite their high penetrative power.
This effect is only noticeable in the rock gamma simulations because of the attenuation of the lower energy gammas in the rock itself and in the water tank and scintillator volumes. The effect of high energy gammas produced in the internal components of the detector would be too low to measure. The flux of high energy gammas from laboratory rock walls will be measured when LZ starts collecting data.


\subsection{Internal Radon}

Radon emanates from detector materials and residual dust in the internal surfaces of the TPC into the liquid bulk. LZ has a requirement of $<$ 2.0~$\mu$Bq/kg of \isotope{Rn}{222}, equivalent to 14 mBq in the active xenon. The current projections conservatively assume emanation at room temperature, whereas emanation from many materials will be decreased at lower temperatures.

One of the daughters of the \isotope{Rn}{222}-chain is \isotope{Bi}{214}, whose 2448~keV gamma line cannot be separated from the \ndbd~ROI by energy resolution alone. However, this gamma line is not a problem for events well-centralized in the xenon as it is vetoed by a coincident $\upbeta$ decay. With the low charge detection threshold of LZ (50\% efficiency at 1.5~keV$_{ee}$) the event will be rejected in $>$99.97\% of decays by coincidence with the $\upbeta$.

A problematic background from radon-induced \isotope{Bi}{214}~is a naked-$\upbeta$ decay, i.e., a $\upbeta$ emission without any accompanying $\upgamma$ emission, with a Q-value of 3.27~MeV and branching ratio of 19.1\%. At 1.0\% energy resolution, 0.5\% of \isotope{Bi}{214}~decays will result in a single-scatter event in the ROI. However, the daughter is \isotope{Po}{214}~which decays by $\upalpha$ emission with a 163.6~$\upmu$s half-life, so can be easily detected and used to veto \isotope{Bi}{214}~decays in the active xenon. 
By excluding all events with either an $\upalpha$ or dead time in the following 2.5~ms, more than 99.99\% of internal \isotope{Bi}{214}~decays will be rejected. This results in around 0.03 background events per tonne in the ROI.


Some of the \isotope{Rn}{222} daughters are positively charged and will be captured on the TPC walls or drift to the cathode, where they can decay and produce a background that cannot be vetoed using a coincident decay. In EXO-200 \cite{Albert:2015vma}, the internal \isotope{Bi}{214}~activity in their fiducial volume was found to be only 11.6\% of the \isotope{Rn}{222} activity. Evidence for this effect has also been seen in LUX \cite{Akerib:2014rda}. The main source of background generated by this population occurs when both the $\upbeta$ particle from \isotope{Bi}{214} and the $\upalpha$ particle from \isotope{Po}{214} are absorbed by the walls or by the cathode grid wires and the 2448~keV $\upgamma$ interacts in the liquid xenon. To study this background, we used the same simulations used to predict the background from the cathode grid with an additional efficiency factor to account for the 25\% of events where both the $\upbeta$ and  $\upalpha$ are absorbed by the walls, as well as the fraction of mass inside the active region. If we assume that all the \isotope{Bi}{214} drifts to the cathode, we predict around 0.41 events in the inner 967~kg volume from this source in 1000 days. This value can be reduced further by the observation of the recoil signal from \isotope{Pb}{210} after the $\upalpha$ emission.


\subsection{Internal \isotope{Xe}{137}}
\label{subsec:Xe137}

Muon-induced and radiogenic neutrons can lead to production of \isotope{Xe}{137} through the reaction \isotope{Xe}{136}(n,$\gamma$)\isotope{Xe}{137}. This isotope is a $\upbeta$-emitter with a Q-value of 4173~keV and a half-life of 3.8 minutes, and therefore this $\upbeta$-decay spectrum overlaps with the \ndbd~ROI. It undergoes a naked $\upbeta$-decay with a branching ratio of 67\%. 

EXO-200, located at the Waste Isolation Pilot Plant (WIPP) in New Mexico (USA), measured 338$^{+132}_{-93}$ \isotope{Xe}{136}(n, $\upgamma$)\isotope{Xe}{137} captures per year by fitting data in coincidence with hits in their muon veto. They calculate that 1.5\% of such captures resulted in a background event in their ROI, or 5.1$^{+2.0}_{-1.4}$ events per year. The fiducial volume for their analysis contained 76.5~kg of \isotope{Xe}{136} \cite{Albert:2015vma}, implying 70~ROI events/(tonne \isotope{Xe}{136})/year \cite{EXO200::2015wtc}. LZ has three advantages that reduce the muon-induced \isotope{Xe}{137} background relative to EXO-200.

\begin{enumerate}
\item The muon flux in the Davis cavern is calculated to be 6.2 $\times$ 10$^{-9}$~cm$^{-2}$ s$^{-1}$ \cite{LZTDR:2016,Akerib:2020ewf}, nearly 100 times lower than at WIPP (4.0 $\times$ 10$^{-7}$~cm$^{-2}$ s$^{-1}$). The calculated flux agrees within 20\% with the measurements carried out with the veto system of the Davis experiment \cite{PhysRevD.27.1444} and the veto system of the Majorana demonstrator \cite{Abgrall:2016cfi} (see \cite{Akerib:2020ewf} for further discussion).
\item Only 8.9\% of natural xenon is \isotope{Xe}{136}. On the other hand, nearly half is \isotope{Xe}{129}~or \isotope{Xe}{131}~, each of which have neutron capture cross sections more than five times larger than \isotope{Xe}{136} \cite{Ni:2007ih}.
\item The LZ detector's large active xenon mass, xenon skin veto, and surrounding outer detector should enable enhanced muon veto efficiency as well as enhanced detection of the neutron capture cascade gamma rays produced in the \isotope{Xe}{136}(n,$\upgamma$)\isotope{Xe}{137} reaction. By looking back over several \isotope{Xe}{137} half-lives for these signatures, potential \isotope{Xe}{137} background may be vetoed as in EXO but with higher efficiency.
\end{enumerate}
The muon flux reduction and \isotope{Xe}{136} abundance alone result in a factor of 100--1000 fewer events in the ROI per kg of xenon compared to EXO-200 \cite{Albert:2015vma}. Furthermore, a muon veto efficiency higher than 99\% can be reached in LZ. Any muon that crosses the outer detector, skin or the TPC of LZ will deposit enough energy to be tagged with close to 100\% efficiency \cite{Akerib:2020ewf}. XENON1T estimates a veto efficiency of 99.5\% for muon tracks in their water tank from Cherenkov light alone \cite{Aprile:2014zvw}. Finally, a neutron capture or multiple scattering event in the LXe active region could also be tagged with high efficiency. Under these conditions, the estimated background from muon-induced neutron capture on \isotope{Xe}{136} is $\ll$ 0.01 events in the ROI and inner 967~kg volume over the 1000 day exposure. The presence of other Xe isotopes in the bulk with higher probability of absorbing incident neutrons further reduces this background by more than an order of magnitude. This background is therefore considered negligible for this search. Other muon-induced backgrounds have also been studied and are similarly negligible.

Radiogenic neutrons may also cause \isotope{Xe}{137} production through thermal neutron capture. Though the thermal neutron flux is very small within the LZ shielding tank, it is higher outside the shielding, measured to be 1.7$\times$10$^{-6}$~cm$^{-2}$ s$^{-1}$ within the Davis cavern \cite{Best:2015}. Xenon within the purification system is exposed to this thermal neutron flux, and given the 4000~kg/day xenon purification rate, 10~kg of xenon is delivered to LZ over the 3.8 minute half-life of \isotope{Xe}{137}, about half of which will go into the TPC. This mechanism produces about 0.28 events in the ROI and inner 967~kg volume over the 1000 day exposure. In addition, some \isotope{Xe}{137} produced by thermal neutron capture in the purification system will decay in the LXe conduit passing through the shielding tank before reaching the LZ time projection chamber.

\subsection{Other Backgrounds}

The two-neutrino decay mode of \isotope{Xe}{136} can result in a background in the ROI as it has the same Q-value as \ndbd. However, the spectrum falls off sharply at the end point. Using the \dbd~spectrum taken from Ref. \cite{Kotila:2012zza}, for 1.0\% energy resolution this results in less than 0.01 background events in 1000 days and in the inner 967~kg volume, or 6.9 $\times$ 10$^{-6}$ events/kg \isotope{Xe}{136}/year in the $\pm~1\sigma$ ROI.

\isotope{B}{8} solar neutrinos can scatter elastically with electrons from the LXe target with enough energy to produce a signal at the Q$_{\beta\beta}$ range. The calculated elastic scattering cross sections per neutrino flavour for charged and neutral current interactions were calculated using the average electron density for Xe atoms and considering a neutrino flux of 5.79 $\times$ 10$^{6}$~cm$^{-2}$ s$^{-1}$ \cite{Bahcall_2005} and the survival probabilities for electron neutrinos of $P_{ee} = 0.543$ \cite{Baudis:2013qla}, resulting in a rate of 0.01 events/tonne/year in the ROI, corresponding to 0.03 events over 1000 days of live time and in the inner 967~kg volume, .

The neutrino capture process $\nu_e$+\isotope{\text{Xe}}{136}$\longrightarrow$e$^-$+\isotope{\text{Cs}}{136} can also induce a background for this search or for an eventual dedicated run with enrichment of \isotope{Xe}{136} in LZ. Estimates from the nEXO experiment predict a rate of 0.3 events/tonne/year, or 0.8 counts/tonne/1000 days with enriched \isotope{Xe}{136} with no single scatter selection \cite{Albert:2017hjq}. In LZ this process can occur on any of the naturally present xenon isotopes, and can produce a background from both the prompt emission of an energetic electron and the subsequent decay of the produced cesium nucleus. The prompt e$^-$ emitted from the neutrino capture in any Xe isotope is expected to have an energy far greater than Q$_{\beta\beta}$ and the prompt de-excitation of the Cs nucleus will likely produce several gammas that can be tagged as multiple scatters with high efficiency \cite{Elliott:2017bui, Albert:2017hjq, Ejiri:2013jda, PhysRevC.99.014320}. The only isotope of cesium that can produce a background in LZ is \isotope{Cs}{136}, since all other isotopes either decay via $\upbeta^+$ emission that can easily be tagged or do not have enough energy to generate a background at Q$_{\beta\beta}$. The decay of \isotope{Cs}{136} into \isotope{Ba}{136} with a half-life of 13.16~days and Q=$2548.2$~keV always produces a cascade of gammas along with the beta emission. This is expected to be completely excluded by multiple scatter rejection, even if the \isotope{Cs}{136} is not removed from the LXe bulk by the purification system. A simulation of 10$^{7}$ decays of \isotope{Cs}{136} in the bulk returned no single scatter events within the ROI. This background is therefore considered negligible for this analysis.



\section{SENSITIVITY PROJECTION}
\label{sec:sensitivity}

The sensitivity to \ndbd~decay is defined as the median 90\% confidence level (CL) upper limit on the number of signal events that would be obtained from a repeated set of background-only experiments, assuming 1000 days of detector live time and a LXe mass of 5.6 tonnes, corresponding to a 1360~kg$\cdot$years exposure of \isotope{Xe}{136}. 
To estimate the background contribution for such an exposure, a multidimensional background model is constructed that accounts for each of the sources discussed in section \ref{sec:backgrounds}. Each background is described by the three observables: energy ($2000 < E < 2700$~keV), depth ($ 2 < z < 132.6$~cm) and radial position ($r < 68.8$~cm). These are combined to model the background with a probability density function (PDF) $P(E, r^2, z)$. For the detector, cavern, and internal radon backgrounds, the PDFs are empirically determined from energy deposit simulations and are approximated by decomposing into factorised energy and position distributions $P(E, r^2, z)=P(E)P(r^2, z)$, which has been verified as a suitable approximation in the given ranges of the observables. The following selection criteria are applied to the simulations to reject background events:

\begin{itemize}
\item Fiducial Volume:  events that occur close to the TPC walls and grids are rejected with a fiducial cut. The extended fiducial volume is defined as 4~cm from the TPC walls, 2~cm above the cathode grid and 13~cm below the gate grid, which defines a region containing 5.6 tonnes of LXe. This cut removes backgrounds which may originate from the grids or TPC walls such as the $\upbeta$-emitting charged \isotope{Rn}{222} daughters.
\item Single Scatter: \isotope{Xe}{136}~\ndbd-decay events in LXe are almost point-like and therefore are expected to produce a single-scatter S2 pulse, whereas the dominant $\upgamma$-ray background predominantly results in extended, multiple scatter events. These are rejected by requiring that multiple vertices are separated by less than 3~mm in the vertical direction, i.e., $\Delta z <3$~mm.  
\item Veto:  $\upgamma$-ray backgrounds that produce a single scatter in the TPC but deposit more than 100~keV either in the outer detector or in the skin vetoes within a narrow 1\,$\upmu$s time window are rejected. This has a significant effect on reducing the background from the \isotope{Tl}{208} 2614.5~keV line, which originates from both the detector components and cavern walls.
\end{itemize}

The remaining backgrounds are expected to have uniform position distributions and are therefore characterised by their energy spectra alone. The \isotope{Xe}{136} \dbd~decay spectrum is from Ref. \cite{Kotila:2012zza} and the \isotope{Xe}{137} $\upbeta$-decay spectrum is obtained from Ref. \cite{betashape}. To model the finite energy resolution, each of the energy distributions are smeared using the LUX energy resolution function \cite{LUXER:2016} that has been scaled to ensure $\sigma/E = 1\%$ at the Q-value.

The \isotope{Xe}{136} \ndbd~signal is modelled with a uniform position distribution and a Gaussian energy distribution centered at Q$_{\beta\beta}$. 
The signal efficiency is estimated to be $80\%$ after simulating signal events with initial kinematics generated using DECAY0 \cite{decay0} and applying the selection criteria. The inefficiency is due to the rejection of multiple scatter signal events arising from Bremsstrahlung emission.

The signal and background PDFs are combined to form the unbinned extended likelihood function,
\begin{align}
    &L(\mu_s, \{\mu_b\}) =\\ \nonumber
    & \left [\mu_s P_s(E, r^2, z) + \sum_{i=1}^{n_b} \mu_b^i P_b^i(E, r^2, z) \right] \prod_{j=1}^{n_b} g(a^j_b, \sigma^j_b),
\label{eq:Likelihood}
\end{align}
where the floating parameters are $\mu_s$, the number of signal events, and $\mu^i_b$, the number of events for the $i$-th background source. The systematic uncertainties $\sigma^j_b$ on the expected background rates $a^j_b$ are included by treating the background sources as nuisance parameters with the set of Gaussian constraint terms $g(a^j_b, \sigma^j_b)$. Table \ref{tab:model} summarises the background sources included as parameters in the likelihood as well as the relative systematic uncertainties on their rate. The cavern background uncertainties are taken directly from the uncertainty of the measured \isotope{U}{238} and \isotope{Th}{232} activities \cite{lab5}. The uncertainties of the detector component background rates are conservatively set at 30\%. The contribution of the remaining background components to the sensitivity is low and their uncertainties are negligible in the first approximation to the final result. The uncertainty for the \isotope{Rn}{222} component is driven by the range of the estimated contamination and is expected to be conservative. Uncertainties for \isotope{Xe}{136} and  \isotope{B}{8} come from the measured half-life and uncertainties of the neutrino flux, respectively, making these rather constrained. Since the \isotope{Xe}{137} background and \isotope{Bi}{214} cathode background rates are not known they are assigned a large uncertainty. The backgrounds considered in this analysis and the associated uncertainties will be measured with high precision once LZ begins collecting data.

The 90$\%$ CL upper limit on the number of signal events is calculated using the profile likelihood ratio (PLR) method, utilising the asymptotic one-sided profile likelihood test statistic \cite{Cowan:2010js}. It has been verified that Wilk's theorem is valid and that the asymptotic approximation is applicable.

The sensitivity analysis takes advantage of the precise multi-parameter reconstruction of events in the LXe TPC, namely the energy and 3-dimensional position, for enhanced sensitivity. As demonstrated by Figure \ref{fig:fiducialization}, the self shielding LXe of LZ results in a low background inner region of the TPC where the majority of signal sensitivity is expected. However, the analysis utilises an extended fiducial volume which allows for the fit of the backgrounds close to the TPC walls and therefore constrains the background in the inner volume of the TPC. Alongside this, the full shape of the position distribution can be used to discriminate between signal-like and background events, which results in both increased signal exposure and sensitivity compared to a simple cut-and-count analysis. Similarly, the extended energy range used in the PDFs strongly constrains the backgrounds near the Q-value. Using an extended phase-space in the profile likelihood analysis improves the sensitivity result by a factor of two when compared to a simple cut-and-count analysis.

\begin{table}[]
 \caption{\label{tab:model}Summary table of the individual background sources and the relative uncertainties on their background rates assumed in this analysis.}
\begin{tabular}{ m{3.5cm}  m{1.5cm} }
 \hline
 Background & $\sigma/N $ \\
 \hline
 \hline
 \vspace{5pt}
 \isotope{U}{238} (Detector) & 30$\%$  \\
 \isotope{Th}{232} (Detector) & 30$\%$  \\
 \isotope{Co}{60} (Detector) & 30$\%$\\
 \isotope{U}{238} (Cavern) &  50$\%$ \\
 \isotope{Th}{232} (Cavern) &  30$\%$ \\
 \isotope{Bi}{214} (Cathode) & 50$\%$\\
 \isotope{Rn}{222} (Internal) &  50$\%$\\
  \isotope{Xe}{137} (Internal) & 50$\%$\\
 \isotope{Xe}{136} \dbd & 5$\%$ \\
 \isotope{B}{8} solar $\nu$ & 5$\%$ \\
  \hline
\end{tabular}
\end{table}



\subsection{Projection with Natural Abundance of \isotope{Xe}{136}}

\begin{figure}[ht]
\begin{center}
\includegraphics[width=\linewidth]{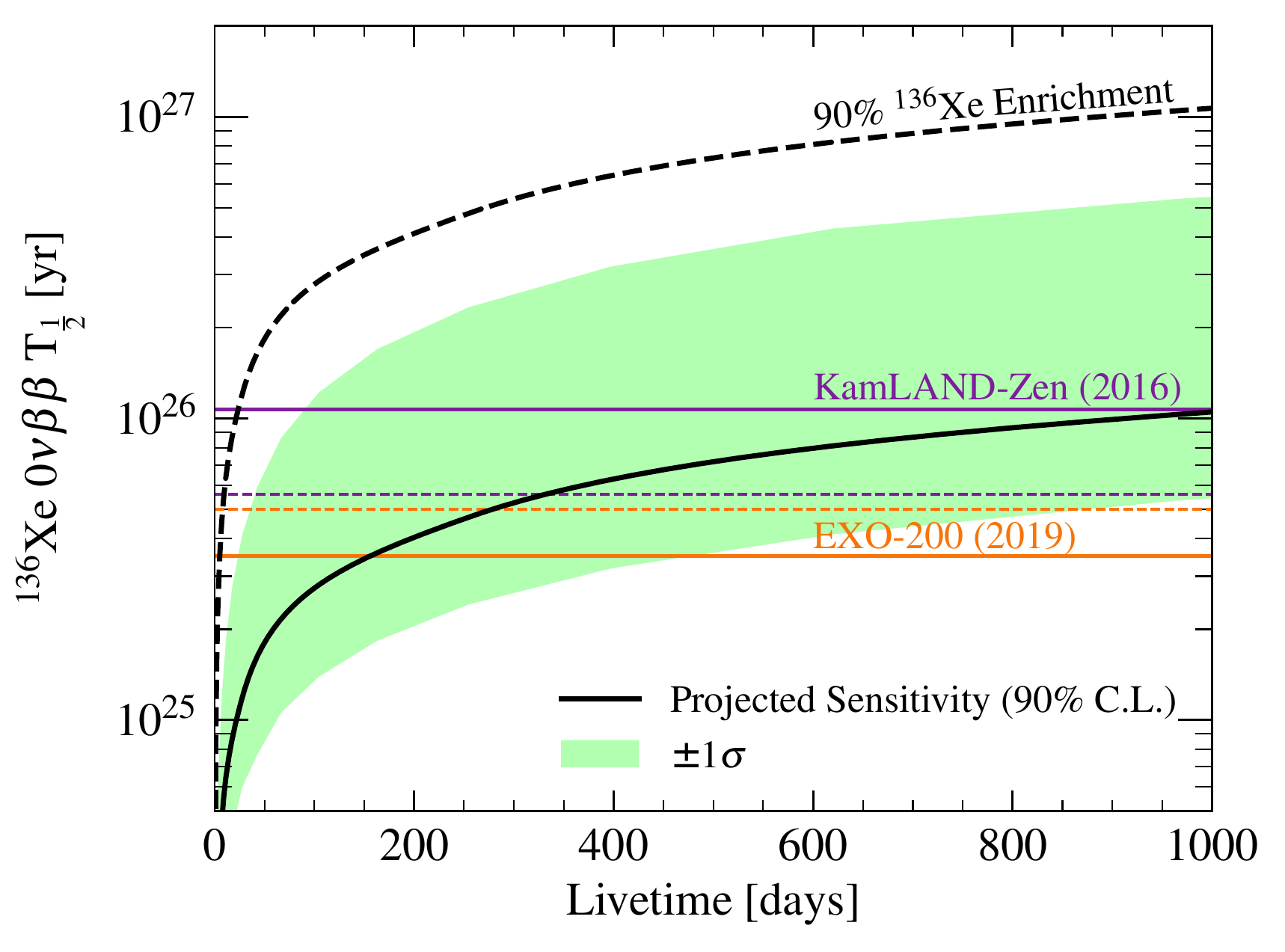}
\caption{LZ projected sensitivity to \isotope{Xe}{136} \ndbd~decay as a function of detector live time. The light green shaded band represents a $\pm$1$\sigma$ statistical uncertainty on the sensitivity. The dashed black line shows the projected sensitivity to \isotope{Xe}{136} \ndbd~decay for a dedicated run with 90\% \isotope{Xe}{136} enrichment. For comparison, the limits set by EXO-200 \cite{Anton:2019wmi} (orange full) and KamLAND-Zen \cite{KamLAND-Zen:2016} (purple full) are also shown, along with the respective projected sensitivities (dashed).}
\label{fig:livetime}
\end{center}
\end{figure}

The 90$\%$ CL sensitivity to the \isotope{Xe}{136 } \ndbd~half-life as a function of detector live time is shown in Fig. \ref{fig:livetime}. A median sensitivity to a half-life of \sensitivityResult~years is reached after 1000 live-days.

\begin{figure}[ht]
\includegraphics[width=\linewidth]{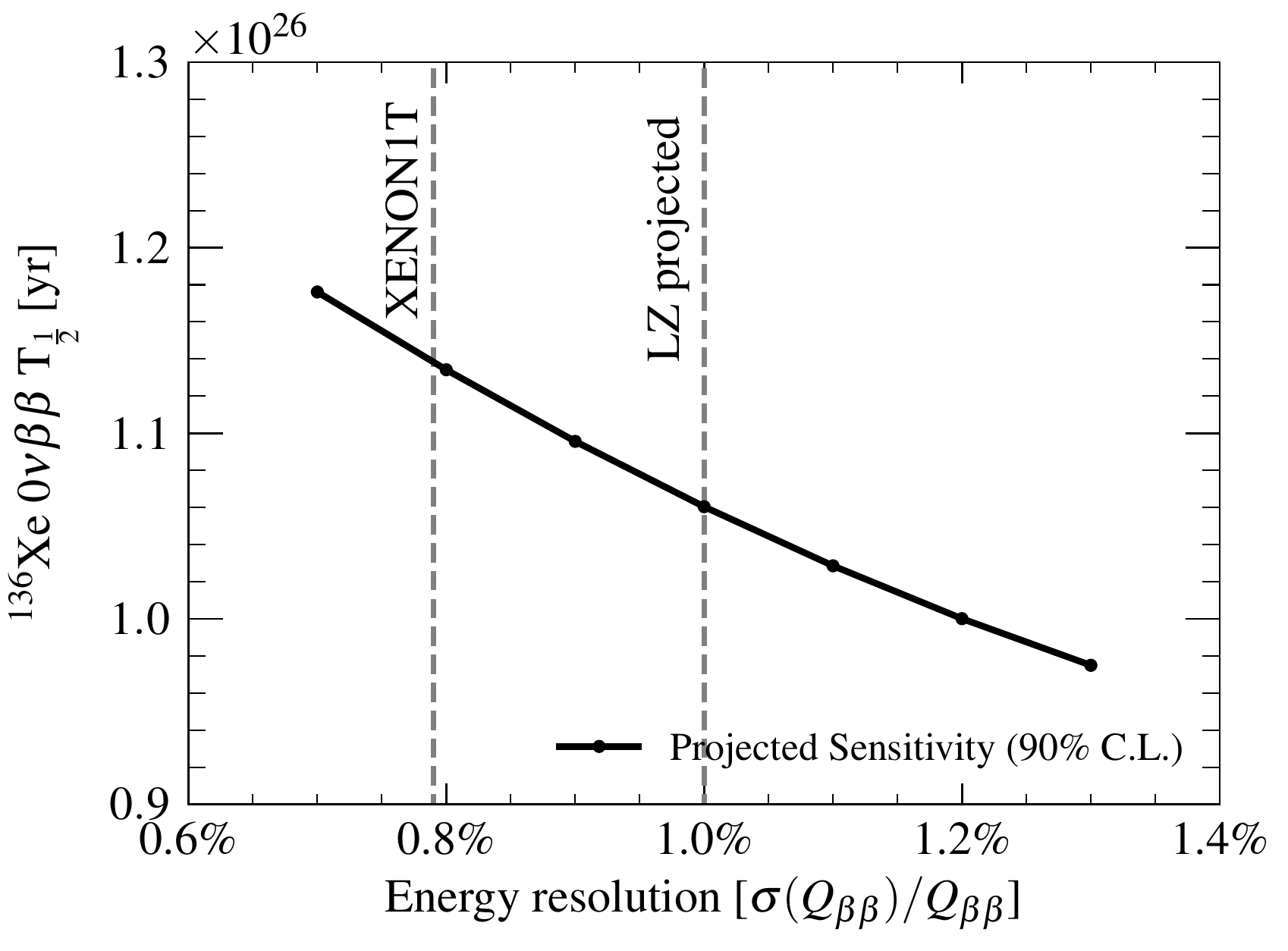}
\caption{\label{fig:EnergyResolution}Expected 90\% CL sensitivity for a 1000 live-days run and for various assumed values of energy resolution at Q$_{\beta\beta}$. The vertical dashed line labeled ``LZ projected'' marks the assumed resolution for this analysis. Also shown on the plot is the projected LZ sensitivity assuming the energy resolution recently measured in XENON1T \cite{Aprile:2020yad}.}
\end{figure}

As the ability to distinguish signal events from the neighbouring \isotope{Bi}{214} and \isotope{Tl}{208} peaks relies heavily on the energy resolution, the dependence of the sensitivity on the energy resolution at the \isotope{Xe}{136} Q-value is shown in Figure \ref{fig:EnergyResolution}. It is clear that an energy resolution slightly worse than the assumed 1.0\% has a minor impact on the sensitivity. However, if the energy resolution were 2.0\% or larger, the impact from the \isotope{Tl}{208} peak would be significant. 

\begin{figure}[ht]
\includegraphics[width=\linewidth]{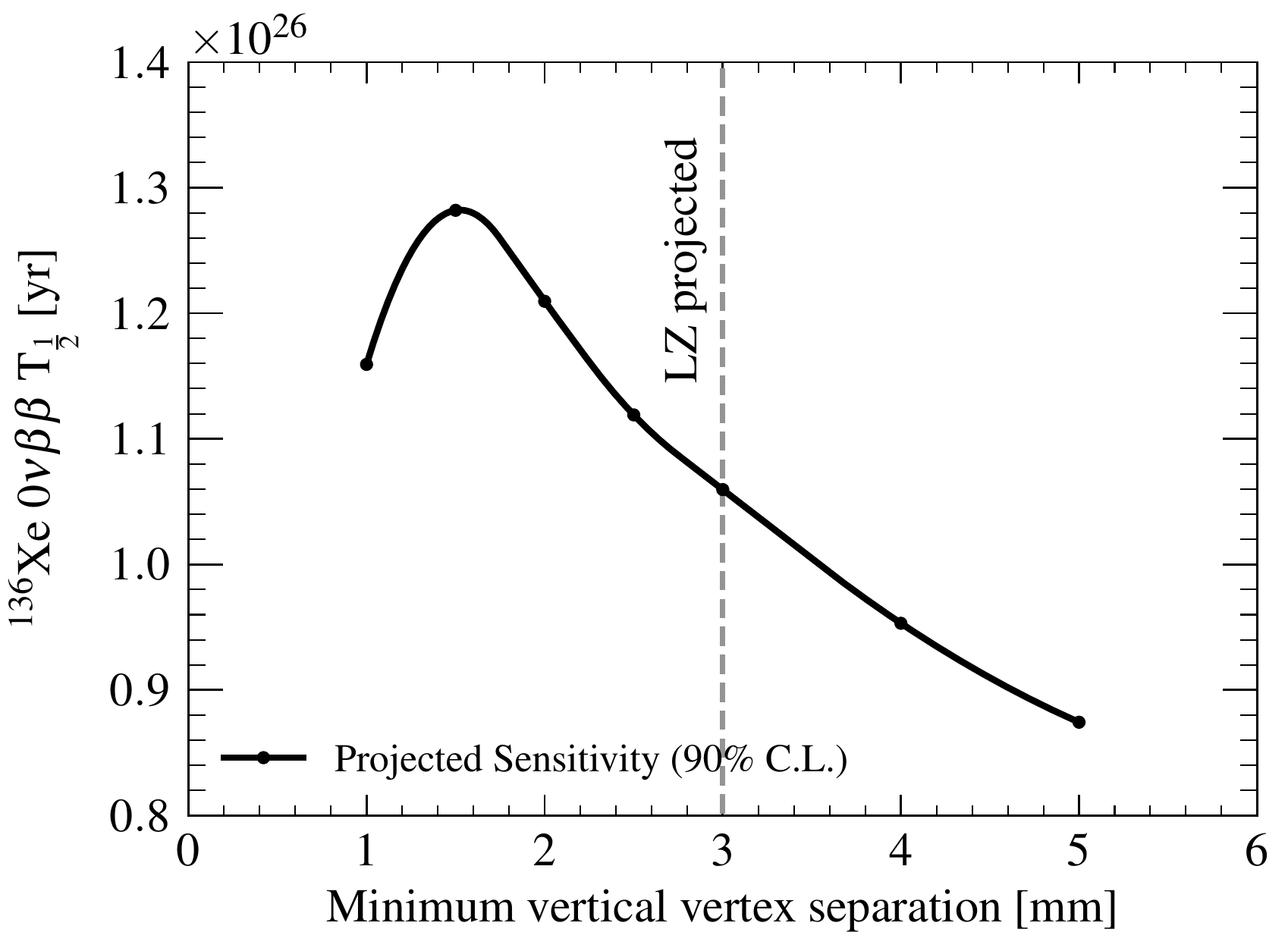}
\caption{\label{fig:dZ}Expected 90\% CL sensitivity for a 1000 live-days run and for various assumed minimum separable vertex distances in depth. Here multiple scatter events are assumed to be rejected based on $z$ separation only. The vertical dashed line marks the assumed separation of 3~mm. At lower separation values, this cut also begins to exclude signal events, resulting in the observed loss in sensitivity.}
\end{figure}

It is assumed in this analysis that multiple scatter events can be rejected with a depth-based vertex separation cut, as multiple energy deposits at different depths in the TPC will have multiple S2 pulses. As expected, Figure \ref{fig:dZ} demonstrates that there is a large variation in sensitivity with this cut as multiple scatter events form the dominant background contribution.

Under the assumption that light neutrino exchange is the driving mechanism for \ndbd, the half-life sensitivity can be translated into the sensitivity to the effective neutrino mass $\left <m_{\beta\beta} \right >$  through the relation \cite{Dolinski:2019nrj}
\begin{equation}
   \left(T_{1/2}^{0\nu}\right)^{-1} = \frac{\left <m_{\beta\beta} \right >^2}{m_{e}^2} G^{0\nu} |M^{0\nu}|^2.
   \label{eq:half-life}
\end{equation}

\begin{figure}[ht]
\begin{center}
\includegraphics[width=\linewidth]{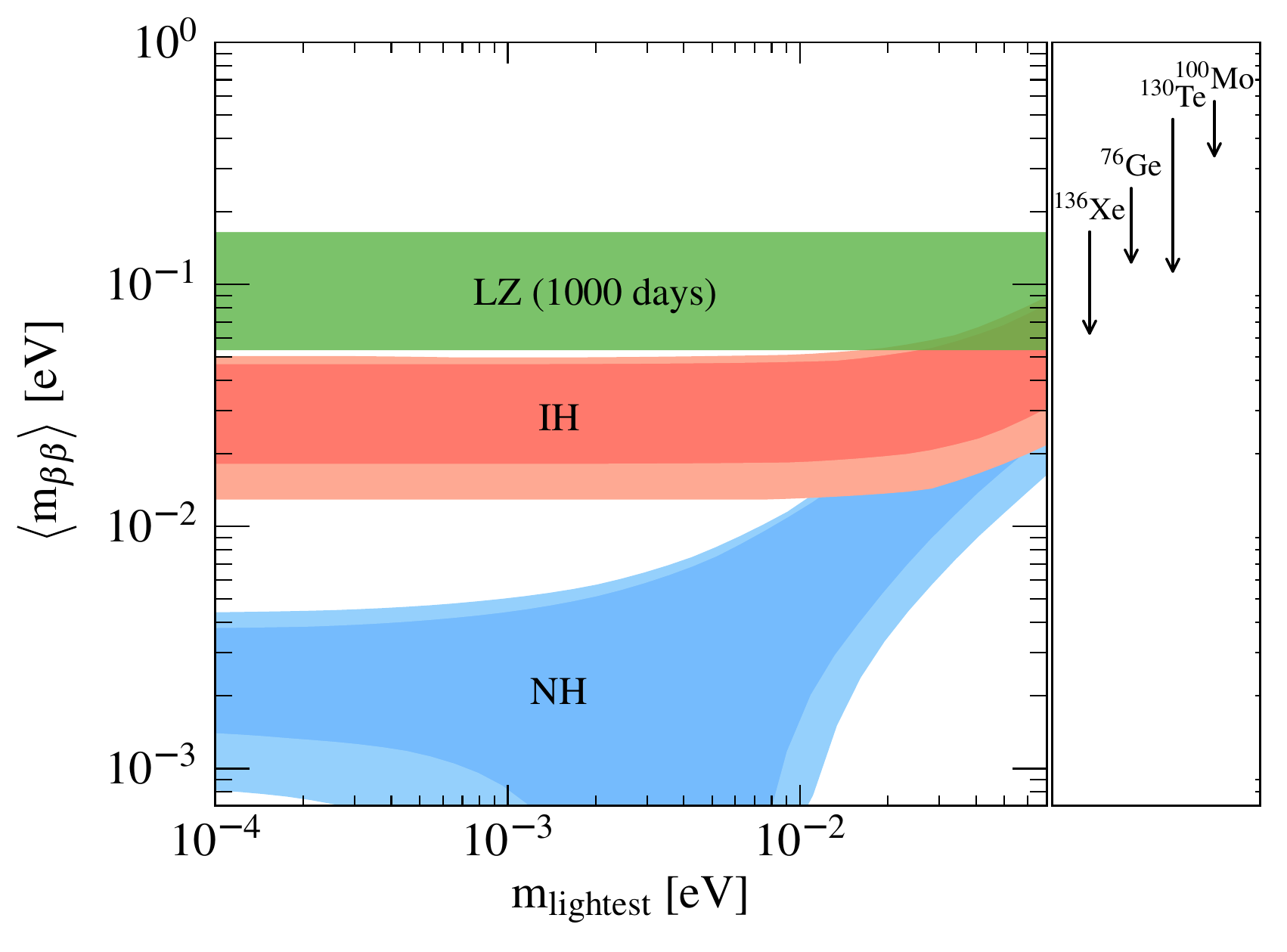}
\caption{LZ projected sensitivity to $\left <m_{\beta\beta} \right >$ and subsequently the neutrino mass hierarchy. The width of the green sensitivity band is due to the uncertainty in the nuclear matrix elements \cite{NME:SkyrmeQRPA}\cite{NME:NREDF}. The red and blue contours show the allowed parameter space ($\pm~1\sigma$) for the inverted hierarchy and normal hierarchy neutrino mass scenarios, respectively \cite{Albert:2017hjq}. On the right are the current best limits and their uncertainties for different \dbd~isotopes, showing that \isotope{Xe}{136} provides the most stringent constraints on $\left <m_{\beta\beta} \right >$ \cite{Dolinski:2019nrj}.}
\label{fig:hierarchy}
\end{center}
\end{figure}

Fig.~\ref{fig:hierarchy} shows that the expected sensitivity to  $\left< m_{\beta\beta} \right>$ after 1000 days is 53--164~meV, with the uncertainty driven by the range of estimates used for the nuclear matrix element \cite{NME:SkyrmeQRPA,NME:NREDF}. The phase space factor from Ref. \cite{Kotila:2012zza} and an unquenched axial-vector coupling constant of $g_{A}$ = 1.27 are used to calculate the effective neutrino mass. 

\subsection{Projection with 90\% \isotope{Xe}{136} Enrichment}

After completion of the WIMP search run of LZ, the sensitivity for \ndbd~could be extended with several specific upgrades that would be either unnecessary or disadvantageous for a dark matter search. The simplest version would be simply to fill the same detector with enriched xenon with no additional upgrades over the WIMP search run. However, with more \isotope{Xe}{136} in the active volume of the detector some backgrounds specific to this isotope would increase and need to be accounted for, namely \isotope{Xe}{137} production from thermal neutron capture when the xenon is in the purification system outside the water tank and the \dbd~decay of \isotope{Xe}{136}. With 1.0\% energy resolution the additional background from \dbd~was verified to be negligible and does not impact the sensitivity. A 10-fold increase in the production of \isotope{Xe}{137} would result in a 9\% decrease in the final sensitivity, but this background can be significantly reduced through the installation of a neutron shield around the xenon purification system. It has been shown that a layer of high-density polyethylene 20~cm thick around the xenon purification tower would reduce the flux of radiogenic neutrons by 2 orders of magnitude \cite{Lemrani:2006dq,TOMASELLO201070}. The impact of shielding the purification system on the production of \isotope{Xe}{137} will be studied in more detail in a dedicated analysis. As discussed in Section \ref{subsec:Xe137}, the capture of muon-induced neutrons by \isotope{Xe}{136} is still considered negligible despite the 90\% abundance and the reduced capture rate on other isotopes. The expected background rate from muon-induced \isotope{Xe}{137} is estimated to be $<$ 0.02 events in the ROI and inner 967~kg volume over the 1000 day exposure, considering a 99\% efficiency at tagging either the muon, the induced neutron or the gamma cascade produced by the neutron capture on \isotope{Xe}{136}. Figure \ref{fig:livetime} demonstrates that with 90\% enriched xenon the sensitivity for a 1000 day run would reach \sensitivityEnriched~years.

\section{CONCLUSIONS}

LZ will be a multi-purpose experiment, capable of exploring a plethora of rare-event phenomena beyond dark matter search due to its ultra-low background environment, great background discrimination potential, large active target mass and an excellent energy resolution.

The LZ experiment will search for \ndbd~decay with a projected median 90\% CL exclusion sensitivity of \sensitivityResult~years for the half-life of \isotope{Xe}{136}, and a sensitivity to $\left<m_{\beta\beta}\right>$ of 53--164~meV. This result assumes 1\% energy resolution at the Q-value of \ndbd~decay and 3~mm vertex separation in depth. The expected background rate within the $\pm$~1$\sigma$ ROI around the Q-value of \ndbd~and in an inner 967~kg fiducial volume is about 35~events in 1\,000~days of live time. The profile likelihood method benefits from background constraints that result from using a larger volume and energy range. 
This sensitivity result demonstrates the potential of a two-phase LXe TPC for searching for \ndbd, as a competitive sensitivity can be reached even for an experiment primarily designed for WIMP detection.

It would be possible, with no improvements in detector parameters and proper mitigation of the \isotope{Xe}{137} neutron-induced background, for LZ to conduct a dedicated post WIMP search run with enriched \isotope{Xe}{136} that would lead to a sensitivity of \sensitivityEnriched~years.


\begin{acknowledgments}

The research supporting this work took place in whole or in part at the Sanford Underground Research Facility (SURF) in Lead, South Dakota. Funding for this work is supported by the U.S. Department of Energy, Office of Science, Office of High Energy Physics under Contract Numbers DE-AC02-05CH11231, DE-SC0020216, DE-SC0012704, DE-SC0010010, DE-AC02-07CH11359, DE-SC0012161, DE-SC0014223, DE-FG02-13ER42020, DE-SC0009999, DE-NA0003180, DE-SC0011702,  DESC0010072, DE-SC0015708, DE-SC0006605, DE-FG02-10ER46709, UW PRJ82AJ, DE-SC0013542, DE-AC02-76SF00515, DE-SC0019066, DE-AC52-07NA27344, \& DOE-SC0012447. This research was also supported by U.S. National Science Foundation (NSF); the U.K. Science \& Technology Facilities Council under award numbers, ST/M003655/1, ST/M003981/1, ST/M003744/1, ST/M003639/1, ST/M003604/1, and ST/M003469/1; Portuguese Foundation for Science and Technology (FCT) under award numbers PTDC/FIS-­PAR/28567/2017; the Institute for Basic Science, Korea (budget numbers IBS-R016-D1); University College London and Lawrence Berkeley National Laboratory thank the U.K. Royal Society for travel funds under the International Exchange Scheme (IE141517). We acknowledge additional support from the Boulby Underground Laboratory in the U.K., the GridPP Collaboration~\cite{faulkner:2006:gridpp,britton:2009:gridpp}, in particular at Imperial College London and additional support by the University College London (UCL) Cosmoparticle Initiative. This research used resources of the National Energy Research Scientific Computing Center, a DOE Office of Science User Facility supported by the Office of Science of the U.S. Department of Energy under Contract No. DE-AC02-05CH11231. The University of Edinburgh is a charitable body, registered in Scotland, with the registration number SC005336. The assistance of SURF and its personnel in providing physical access and general logistical and technical support is acknowledged.

\end{acknowledgments}

\bibliography{bibliography}{}
\bibstyle{plain}

\end{document}